\documentclass[prb,aps,twocolumn]{revtex4}
\usepackage{amsmath}
\usepackage{amssymb}
\usepackage{verbatim}
\usepackage{appendix}
\usepackage{graphicx}

\newcommand{\Eq}[1]{Eq.\,(\ref{#1})}% \Eq{abc}
\newcommand{\Tab}[1]{Table\,\ref{#1}}% \Tab{tab:abc}
\newcommand{\Cite}[1]{~[\onlinecite{#1}]} % \Onlinecite{abc}

\DeclareMathAlphabet\mathbfcal{OMS}{cmsy}{b}{n}

\begin{document}
\title{Dipolar polaritons in microcavity-embedded coupled\\quantum wells in electric and magnetic fields}
\author{J. Wilkes}
\author{E. A. Muljarov}
\affiliation{School of Physics and Astronomy, Cardiff University, The Parade, Cardiff CF24 3AA, United Kingdom}

\begin{abstract}
We present a precise calculation of spatially-indirect exciton states in semiconductor coupled quantum wells and polaritons formed from their coupling to the optical mode of a microcavity. We include the presence of electric and magnetic fields applied perpendicular to the quantum well plane. Our model predicts the existence of polaritons which are in the strong coupling regime and at the same time possess a large static dipole moment. We demonstrate, in particular, that a magnetic field can compensate for the reduction in light-matter coupling that occurs when an electric field impresses a dipole moment on the polariton.
\end{abstract}

\maketitle

\section{Introduction}
\label{introSec}

Exciton-polaritons in microcavities in the strong coupling regime have been studied intensively in recent years\Cite{CarusottoRMP2013}. They are realized by placing semiconductor quantum wells (QWs) at the antinode position of a resonant electromagnetic field inside a cavity. Notable advances include the observation of phenomena such as polariton Bose-Einstein condensation\Cite{KasprzakNAT2006,BaliliSCI2007}, superfluidity\Cite{AmoNAT2009} and the formation of quantum vortices\Cite{LagoudakisNATP2008}. A number of applications were developed, such as electrically pumped room temperature polariton lasers\Cite{ChristopoulosPRL2007,BhattacharyaPRL2014} and optical logic devices\Cite{BallariniNATC2013}.

Spatially indirect excitons, formed from an electron and a hole separated into adjacent coupled quantum wells (CQWs) by an external bias are also an attractive system for study\Cite{ButovNAT2002,ButovJETP2016}.
This is due to the possibility to control the exciton properties by varying the applied electric field\Cite{ButovPRB1999}. Macroscopic charge separation causes indirect excitons to have a built-in static dipole moment. This enhances the exciton-exciton interaction strength leading to a luminescence blue shift\Cite{NegoitaPRB2000}. The blue shift provides a probe of the exciton density and was used to study exciton phase transitions\Cite{LaikhtmanEPL2009,Cohen_arXiv2015}. The static dipole moment also facilitates electrostatic\Cite{WinbowPRL2011,LeonardAPL2012,DorowAPL2016,SchinnerPRL2013} and optical\Cite{AlloingSREP2013,HammackPRL2006} control of exciton transport and was exploited for the development of excitonic devices\Cite{AndreakouAPL2014}.

Recently, CQWs were embedded in a planar Bragg-mirror microcavity to create polaritons with a large dipole moment typical for indirect excitons\Cite{CristofoliniSCI2012}. Usually, an indirect exciton is only weakly coupled to light due to the reduced overlap of the electron and hole wave functions. However, when asymmetric CQWs are used, the electric field can be chosen such that the indirect exciton, direct exciton and the cavity mode are brought into resonance, resulting in a strongly coupled three-level system. Such hybrid quasiparticles, called {\em dipolaritons}, present a system with a greater flexibility of control and new possible applications compared to regular polaritons formed from direct excitons. Proposed applications include superradiant THz emission\Cite{KyriienkoPRL2013}, tunable single-photon emission\Cite{KyriienkoPRA2014} and optical parametric oscillators\Cite{KhadzhiJETP2015}. Dipolaritons were also realized in wide single QWs embedded in a dielectric waveguide\Cite{Rosenberg_arXiv2016}. In that work, excitons under bias acquire a static dipole moment because the large QW width permits substantial charge separation. At the same time, excitons remain strongly coupled to light due to the absence of a potential barrier separating electrons and holes, so that their wave functions can have a sufficient overlap.

Applying a magnetic field offers an extra degree of control over excitonic systems as it modifies the properties of excitons and consequently, those of exciton-polaritons. Experiments have demonstrated continuous tuning from the weak to strong coupling regime and control of the vacuum Rabi splitting by varying magnetic field\Cite{TignonPRL1995,FisherPRB1996,PietkaPRB2015}. A magnetic field can suppress the polariton relaxation bottleneck as it enhances the polariton-phonon scattering rates\Cite{BhattacharyaAPL2012}. Magnetic field dependence of the polariton lasing threshold was also investigated\Cite{Kochereshko_arXiv2013}. However, all these works focused on polaritons without a static dipole moment. {\em Magneto-dipolaritons} present an intriguing new system which, to the best of our knowledge, was not yet studied. In addition to the possibility to fabricate structures with different optical resonances and different QW and barrier widths, the combination of both electric and magnetic fields as controllable parameters realizes a highly non-trivial system with a rich physics to explore. This opens an avenue for new theoretical and experimental investigations.

The subject of this paper is a rigorous calculation of exciton-polariton states in microcavity-embedded CQWs and an examination of their dependence on electric and magnetic fields. We show that the magnetic field enhances the exciton-photon coupling and can effectively compensate for the darkening associated with the direct-indirect exciton crossover. This allows polaritons in the strongly coupled regime to have a large static dipole moment that is comparable to the center-to-center distance of the CQWs. We use the multi-sub-level approach (MSLA) recently introduced in Ref.\Cite{SivalertpornPRB2012}. The main benefit of this approach is that we solve precisely the exciton Schr{\" o}dinger equation in three dimensions. In Ref.\Cite{WilkesNJP2016}, we used this method to calculate the electric and magnetic field dependence of the lifetime, Bohr radius, dipole moment, binding energy and effective mass of CQW excitons. We found that these properties strongly depend on the direct-indirect exciton crossover which occurs for changing electric and/or magnetic field. Such a dependence cannot be captured by any two-dimensional (2D) approximations of the exciton that were used by many authors, see e.g. Refs.\Cite{LozovikJETP1997,ArseevJETP1998}. The same method was used to describe the effect of barrier width on the CQW excitons in an electric field\Cite{SivalertpornPLA2016}. In Ref.\Cite{SivalertpornPRL2015}, the MSLA was extended to describe dipolaritons. However, this was done without including the influence of an applied magnetic field, which is the main focus of the present paper.

In Section\,\ref{modelSec}, we describe our method of solving the coupled Schr{\" o}dinger and Maxwell's wave equations used to model polaritons in microcavity-embedded CQWs. Results from the application of our model to a realistic structure are presented in Section\,\ref{resultsSec}. Summary and conclusions are in Section\,\ref{summarySec}. Appendices A-C contain derivations of the results used and provide details of the numerical procedure.

\section{Coupled Maxwell's and material equations}
\label{modelSec}

A microscopic approach to light-matter interaction is used to describe the coupling of QW excitons and cavity photons. The problem entails solving Maxwell's wave equation for the light field $\bf E$, given by
\begin{equation}
\nabla^2 {\bf E} = \frac{1}{c^2}\frac{\partial^2}{\partial t^2} \left( \epsilon_b {\bf E} + 4 \pi {\bf P} \right),
\label{basicWave}
\end{equation}
where ${\bf P}$ is the exciton macroscopic polarization and $\epsilon_b(z)$ is the low-frequency background dielectric constant of the heterostructure. We split out the time dependence of the fields,  ${\bf E}={\bf E}_\omega({\bf R},z) e^{-i \omega t}$ and ${\bf P}={\bf P}_\omega({\bf R},z) e^{-i \omega t}$, and for the planar heterostructures considered here, separate the coordinates that are perpendicular and parallel to the QW plane ($z$ and $\bf R$, respectively), using the factorization
\begin{equation}
{\bf E}_\omega({\bf R},z) = \hat{\bf e}\,\mathcal{E}(z)e^{i{\bf K}\cdot {\bf R}}.
\label{lightField}
\end{equation}
Here, $\bf K$ is the in-plane wave vector and $\omega$ is the frequency of light. For s-polarized light, the unit vector of the light polarization $\hat{\bf e}$ is normal to the growth axis. The macroscopic polarization ${\bf P}_\omega$ is linked via an equation
\begin{equation}
{\bf P}_\omega({\bf R},z) = \int Y({\bf R},\boldsymbol\rho,z_e,z_h) \mathbfcal{M}({\bf r})\,d{\bf r}
\label{macroPolarization}
\end{equation}
to the microscopic exciton polarization $Y$ which in turn satisfies an inhomogeneous Schr\"odinger equation\Cite{Stahl87}
\begin{equation}
\left(\hat{H} - \hbar\omega - i\gamma\right)Y({\bf R},\boldsymbol\rho,z_e,z_h) = \mathbfcal{M}({\bf r})\cdot {\bf E}_\omega({\bf R},z)\,.
\label{materialEqn}
\end{equation}
Here, $\hat{H}$ is the full Hamiltonian of a CQW exciton in external fields~\Cite{WilkesNJP2016}, $\mathbfcal{M}({\bf r})$ is the optical transition dipole moment, ${\bf r}=(\boldsymbol\rho,z_e-z_h)$ is the electron-hole relative coordinate, with $\boldsymbol\rho$ being that in the plane of the QW, and $z_{e(h)}$ is the electron (hole) coordinate in the growth direction. The transition dipole moment is assumed to be isotropic in the QW plane and its magnitude to have the form ${\cal M}({\bf r})= \mu_{\rm cv}\delta({\bf r})$ in the point dipole approximation so that $z = z_e = z_h$ in the right hand side of \Eq{materialEqn}. Here, $\mu_{\rm cv} = ed_{\rm cv}$ is the conduction to valence band dipole matrix element. A phenomenological damping constant $\gamma$ is used to describe non-radiative losses.

Introducing the excitonic non-local susceptibility $\chi(z,z')$ of the CQW structure, defined by
\begin{equation}
{\bf P}_\omega({\bf R},z) = \int \chi(z,z')  {\bf E}_\omega({\bf R},z') dz'\,,
\label{susc}
\end{equation}
we arrive at the integro-differential wave equation for the amplitude $\mathcal{E}(z)$ of the light field:
\begin{equation}
\left(K^2 - \frac{\partial^2}{\partial z^2} \right) \mathcal{E} (z) = \frac{\omega^2}{c^2}\left[ \epsilon_b(z) \mathcal{E} (z) + 4 \pi \int \chi(z,z') \mathcal{E}(z') \, dz' \right],
\label{wave}
\end{equation}
where $K = |{\bf K}|$. The excitonic susceptibility $\chi(z,z')$ is found by solving \Eq{materialEqn} for $Y$, with the help of the spectral representation of the Green's function, as detailed in Appendix~\ref{App0}. It is then expressed as a sum over all quantized exciton states
\begin{equation}
\chi(z,z') = \mu_{\rm cv}^2\hbar\omega\sum_{\nu} \frac{\varphi_{\nu{\bf K}}(0,z,z) \varphi_{\nu{\bf K}}^\ast(0,z',z')}{(E_{\nu{\bf K}}- i \gamma)(E_{\nu{\bf K}}- i \gamma - \hbar\omega)},
\label{susceptibility}
\end{equation}
in which $ E_{\nu{\bf K}}$ is the eigen energy and $\varphi_{\nu{\bf K}}(\boldsymbol\rho, z_e,z_h)$ the wave function describing the internal structure of exciton state $\nu$ with momentum ${\bf K}$. They satisfy a Schr\"odinger equation $\hat{H}_{\rm x}^{\bf K}\varphi_{\nu{\bf K}}=E_{\nu{\bf K}}\varphi_{\nu{\bf K}}$, in which $\hat{H}_{\rm x}^{\bf K}(\boldsymbol\rho, z_e,z_h)$ is the exciton reduced Hamiltonian obtained from the full Hamiltonian $\hat{H}({\bf r}_e,{\bf r}_h)$ by making a unitary transformation to split out the exciton relative and center of mass motion\Cite{LozovikJETP1997}. Here, ${\bf r}_{e,h}$ are the 3D electron and hole coordinates. For the present case of external static electric and magnetic fields applied perpendicular to the CQW structure, this transformation is done by a factorization
\begin{equation}
\Psi_{\nu{\bf K}}({\bf r}_e,{\bf r}_h) = \exp \left(i\left[{\bf K} + \frac{e}{\hbar c}{\bf A}(\boldsymbol\rho) \right] \cdot {\bf R}\right)\varphi_{\nu{\bf K}}(\boldsymbol\rho, z_e,z_h)
\end{equation}
of the full exciton wave function satisfying the full Schr\"odinger equation $\hat{H}\Psi_{\nu{\bf K}}=E_{\nu{\bf K}}\Psi_{\nu{\bf K}}$. Here, ${\bf A}$ is the vector potential of magnetic field ${\bf B}$. Since ${\bf B}$ is applied along the growth direction, the symmetric gauge ${\bf A}({\boldsymbol\rho}) = \frac{1}{2}{\bf B}\times\boldsymbol\rho$ can be used.

In the effective mass approximation, the reduced Hamiltonian for an exciton in static electric and magnetic fields applied perpendicular to the QW plane has the form\Cite{WilkesNJP2016}
\begin{equation}
\hat{H}^{\bf K}_{\rm x}(\boldsymbol\rho,z_e,z_h) = \hat{H}^\perp_e(z_e) + \hat{H}^\perp_h(z_h) + \hat{W}^{\bf K}_B(\boldsymbol\rho) + E_g
+ V_C(r)\,,
\label{Exc-Ham}
\end{equation}
in which
\begin{eqnarray}
\hat{H}^\perp_{e,h}(z) &=& - \frac{\hbar^2}{2}\frac{\partial}{\partial z}\frac{1}{m^{\perp}_{e,h}(z)}\frac{\partial}{\partial z} + U_{e,h}(z)\,,
\label{H_perp}
\\
\hat{W}^{{\bf K}}_B (\boldsymbol\rho) &=& -\frac{\hbar^2}{2 \mu} \nabla^2 - \frac{i e \hbar}{\varkappa c}{\bf A}(\boldsymbol\rho) \cdot \nabla + \frac{e^2}{2\mu c^2}A^2(\boldsymbol\rho)
\nonumber\\
&&+ \frac{K^2}{2M}+ \frac{2e}{Mc} {\bf K} \cdot {\bf A}(\boldsymbol\rho)\,,
\end{eqnarray}
$E_g$ is the band gap, and $V_C(r)=-e^2/(\epsilon_{\rm QW} r)$ is the Coulomb interaction, with the dielectric constant of the QW layers $\epsilon_{\rm QW}$ and $r=|{\bf r}| = \sqrt{\rho^2 + (z_e - z_h)^2}$. Here $M = m^{||}_{e} + m^{||}_{h}$, $1/\mu = 1/m^{||}_{e} + 1/m^{||}_{h}$, and $1/\varkappa = 1/m^{||}_{e} - 1/m^{||}_{h}$, in which we neglect any $z$-dependence of the in-plane electron and hole effective masses $m^{||}_{e,h}$. This is justified by low mass contrast in the heterostructures treated here and a minor contribution of the electron and hole wave functions outside the well regions. For the band gap $E_g$ and in-plane masses $m^{||}_{e,h}$, the values for the QW layers are used. The confinement and the external bias are included in the potentials $U_{e,h}(z) = V^{\rm QW}_{e,h}(z) \pm eFz$ where $F$ is the electric field. The perpendicular masses $m^{\perp}_{e,h}(z)$ are layer-dependent step functions, as are the QW confinement potentials $V^{\rm QW}_{e,h}(z)$.

To obtain the eigenstates contributing to \Eq{susceptibility}, we employ the MSLA developed in Refs.\,\Cite{SivalertpornPRB2012,WilkesNJP2016}. Owing to the high exciton mass $M$ (which further increases in an applied magnetic field\Cite{WilkesNJP2016}) and small values of the photon momentum ${\bf K}$ involved, any ${\bf K}$-dependence of the exciton states can be neglected in the polariton problem treated here. In the MSLA, the wave function describing the internal structure of the exciton, i.e. the solution of the Hamiltonian \Eq{Exc-Ham}, is expanded into the basis of Coulomb-uncorrelated electron-hole pair states,
\begin{equation}
\varphi_{\nu 0}(\boldsymbol\rho,z_e,z_h) = \sum_{n} \Phi_n(z_e,z_h) {\phi}^{(\nu)}_n(\rho) e^{im\phi}\,,
\label{solution}
\end{equation}
where we have used the polar coordinates $\boldsymbol\rho=(\rho,\phi)$ and introduced the exciton angular momentum $m$. Note that only optically active states with $m=0$ contribute to the polariton problem. Each basis function $\Phi_n(z_e,z_h)$ is the product of single-particle electron and hole wave functions which themselves are eigenstates of the perpendicular motion Hamiltonians \Eq{H_perp}. The radial components of the wave function ${\phi}^{(\nu)}_n(\rho)$ are calculated using a matrix generalization of the shooting method with Numerov's algorithm incorporated into the finite difference scheme. Full details can be found in Ref.\Cite{WilkesNJP2016}.

The excitonic susceptibility then takes the form
\begin{equation}
\chi(z,z') = \sum_{n,m=1}^N\Phi_n(z,z)\Phi_m(z',z')\chi_{nm}
\label{chi-kernel}
\end{equation}
with a matrix
\begin{equation}
\chi_{nm} = \hbar\omega\mu_{\rm cv}^2\sum_{\nu=1}^{N_{\rm ex}}\frac{\phi^{(\nu)}_n(0)\phi^{(\nu)}_m(0)}{(E_{\nu0} - i\gamma)(E_{\nu0} - i\gamma - \hbar\omega)}\,.
\label{chinm}
\end{equation}
In the practical implementation of the method, we first solve one-dimensional Schr{\" o}dinger equations for the electron and hole with the Hamiltonians \Eq{H_perp} and find a set of $N$ basis functions $\Phi_n(z_e,z_h)$  determining the size of the matrix $\chi_{nm}$. We then solve the exciton Schr{\" o}dinger equation in external electric and magnetic fields, finding $N_{\rm ex}$ lowest states, in this way calculating $\phi^{(\nu)}_n(\rho)$ and the exciton energies $E_{\nu0}$ determining the matrix $\chi_{nm}$. Typically, $N$ ranges between 4 and 16 and $N_{\rm ex}$ between 10 and 200, depending on the external field values.  Next, for the known excitonic susceptibility $\chi(z,z')$, given in the form of a factorizable kernel \Eq{chi-kernel}, the wave equation \Eq{wave} allows a semi-analytic exact solution, in the form of exponentials with amplitudes determined by the Fourier transforms of the basis functions $\Phi_n(z,z)$ as detailed in Appendix~\ref{appA}. In particular, we determine a $2\times 2$ scattering matrix of the active layer containing the CQWs, which consists of the transmission and reflection coefficients for left and right propagating waves. The scattering matrix of the optically active layer is then incorporated into the standard procedure of the scattering matrix calculation\Cite{KoPRB1988}, giving the optical response of the full system of CQWs embedded in the microcavity.

\section{Application to I\MakeLowercase{n}G\MakeLowercase{a}A\MakeLowercase{s} microcavity-embedded asymmetric CQW\MakeLowercase{s}}
\label{resultsSec}

\begin{table}
\begin{center}
\caption{Parameters of the model}
\begin{tabular}{l l l}
\hline
$\epsilon_{\rm QW}$ & Permittivity in QWs                     & 12.0          \\
$E_g$               & Band gap of $\rm In_{0.1}Ga_{0.9}As$ QW & 1.373\,eV     \\
$m^{\perp}_e(z)$    & Electron mass in QW1                    & 0.0622\,$m_0$ \\
                    & Electron mass in QW2                    & 0.0630\,$m_0$ \\
                    & Electron mass in barrier                & 0.0665\,$m_0$ \\
$m^{\perp}_h(z)$    & Hole mass in QWs                        & 0.338\,$m_0$  \\
                    & Hole mass in barrier                    & 0.34\,$m_0$   \\
$M_{\rm x}$         & In-plane exciton mass                   & 0.22\,$m_0$   \\
$\mu$               & In-plane reduced exciton mass           & 0.040\,$m_0$  \\
$d_{\rm cv}$        & Dipole matrix element                   & 0.6\,nm       \\
$\gamma$            & Damping constant                        & 0.2\,meV      \\
\hline
\label{parameters}
\end{tabular}
\end{center}
\end{table}
\begin{figure}
\centering
\includegraphics[width=0.45\textwidth]{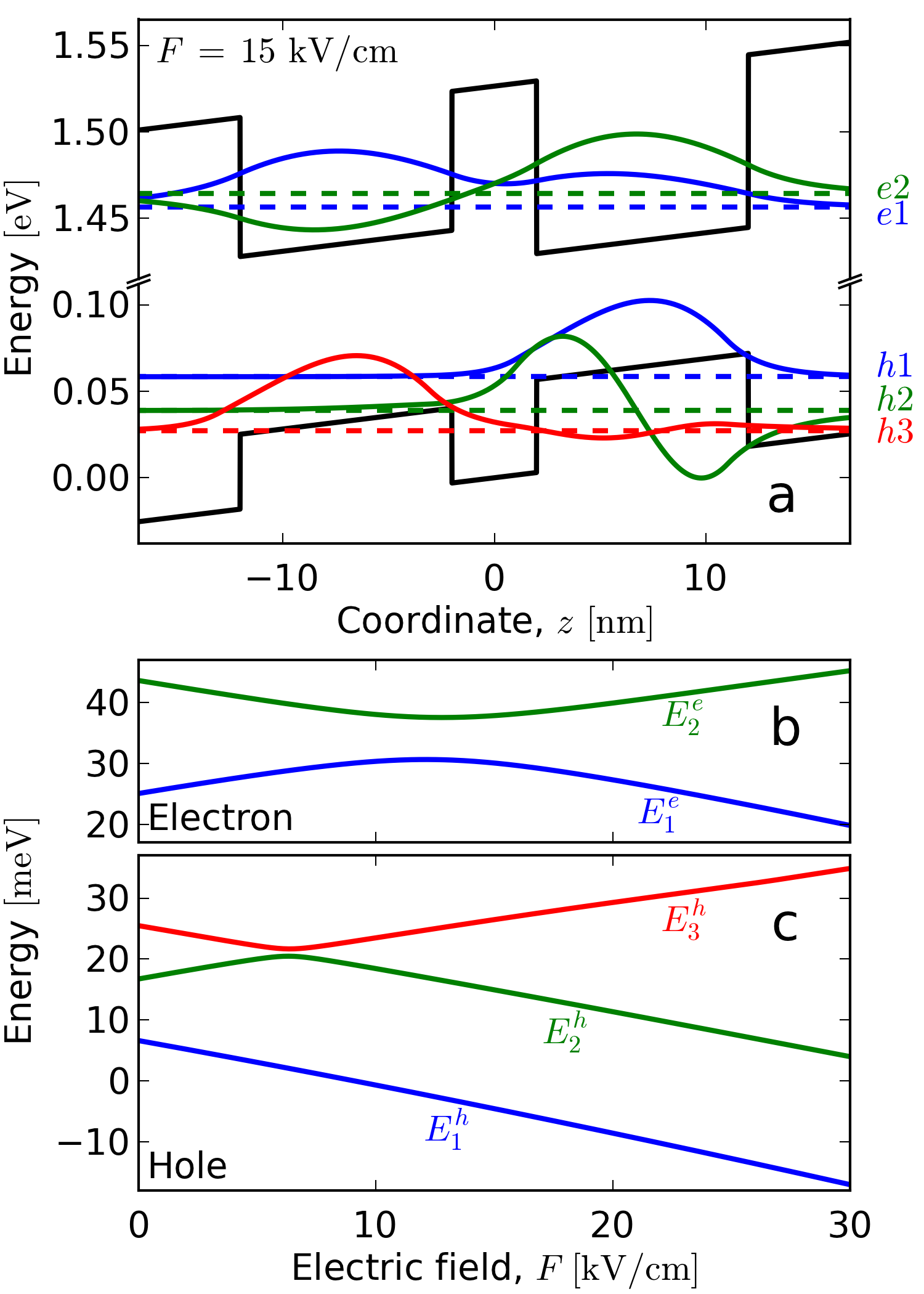}
\caption{(a) Electron band structure and  first few electron ($e1$ and $e2$) and hole ($h1$, $h2$ and $h3$) states for $F = 15\,{\rm kV/cm}$ in 10--4--10-nm InGaAs asymmetric CQWs used in Ref.\Cite{CristofoliniSCI2012}. (b) Electron and hole energy levels as functions of electric field.}
\label{zWfn}
\end{figure}
We consider InGaAs asymmetric CQWs inside a microcavity that were studied in Ref.\Cite{CristofoliniSCI2012}. We note that our approach can also be used to model the recently observed waveguide dipolaritons\Cite{Rosenberg_arXiv2016}. This is the subject of future work and not explored in detail here. \Tab{parameters} gives a list of the parameters used. Fig.\,\ref{zWfn}a shows the electron band structure of the CQWs along the growth axis. The left and right ${\rm In}_x{\rm Ga}_{1-x}{\rm As}$ QWs have $x = 0.08$ and $x = 0.1$, respectively. The confinement potentials are tilted by the perpendicular electric field, $F=15\,{\rm kV/cm}$. The single particle wave functions of the first two electron ($e1$,$e2$) and first three hole ($h1$,$h2$,$h3$) states are also shown, each offset by their energy. These are eigenstates of the Hamiltonian \Eq{H_perp} which neglects the electron-hole Coulomb interaction. They are used to prepare the basis functions, $\Phi_n(z_e,z_h)$, which are required for the expansion of the exciton wave function given by \Eq{solution}. For the range of electric fields considered in this work, sufficient accuracy was achieved using just two electron and three hole states. This results in only $N = 6$ pair states needed for the summation in \Eq{solution}.

At high electric field, tilting of the CQW confinement potential spatially separates the electron and hole ground states ($e1$ and $h1$). The exciton thus acquires a large dipole moment that is comparable to the nominal center-to-center distance between the QWs. The electric field dependence of the electron and hole energies are shown in Figs.\,\ref{zWfn}b and \ref{zWfn}c, respectively. For electrons, one can see that the QW asymmetry is partly compensated at around $12\,{\rm kV/cm}$, forming resonantly tunnel coupled symmetric and antisymmetric states. A similar effect takes place around $6\,{\rm kV/cm}$ for the hole when the ground state in the left QW is resonantly coupled to the first excited state in the right QW.  Away from the tunneling resonance, the derivative of the energy of each single particle state with respect to electric field is approximately equal to the product of charge and the expectation value of the particle's position, $\pm e \langle z\rangle$, and the main features of the system are described by $e1$, $e2$, $h1$ and $h2$. However, the state $h3$ is required for an accurate description around $6\,{\rm kV/cm}$ where $h2$ and $h3$ are close in energy.

\begin{figure}
\centering
\includegraphics[width=0.45\textwidth]{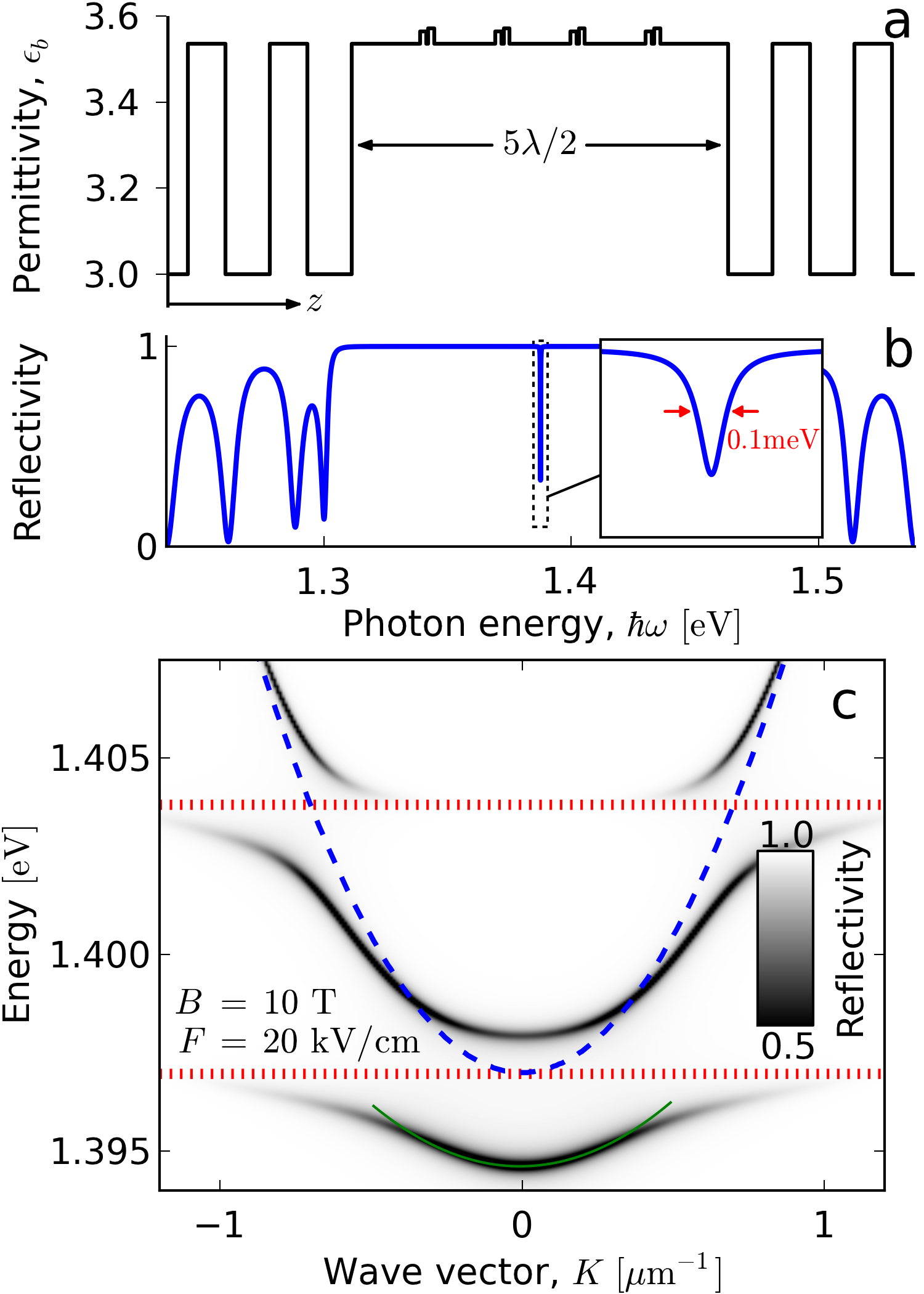}
\caption{(a) The spatial profile of the permittivity in the considered sample, having four pairs of InGaAs CQWs located at the antinode positions of the resonant cavity mode in a $5\lambda/2$ cavity surrounded by  two DBRs formed from 17 and 21 pairs of GaAs/InGaAs layers. (b) Reflectivity spectrum of the cavity in the absence of light-matter coupling. The inset shows a magnification of the cavity mode. (c) In-plane wave vector dependence of the reflectivity spectrum. The bare cavity mode and the exciton dispersion are shown by the dashed blue and dotted red lines, respectively. The solid green line is a parabolic fit to the bottom of the polariton ground state dispersion.}
\label{schematic}
\end{figure}
Fig.\,\ref{schematic}a shows the permittivity of the treated microcavity structure as a function of coordinate $z$. The microcavity consists of 17 and 21 pairs of alternating GaAs and InGaAs $\lambda/4$ layers forming the distributed Bragg reflectors (DBRs). Four pairs of asymmetric CQWs are placed at the anti-node positions of the resonant optical mode in a $5\lambda/2$ cavity. The reflectivity spectrum of the bare cavity (without CQWs) is shown in Fig.\,\ref{schematic}b. A dip in reflectivity occurs at the cavity mode, shown in detail by the inset in Fig.\,\ref{schematic}b. The cavity mode has a full width at half maximum of $0.1\,{\rm meV}$ giving a $Q$-factor of $\approx 14000$.

An example of the full reflectivity spectrum including the light-matter interaction in the CQWs is shown by the grayscale in Fig.\,\ref{schematic}c as a function of in-plane wave vector $K=|{\bf K}$ for $F=20\,{\rm kV/cm}$ and $B=10\,{\rm T}$. The dips in reflectivity are the lowest three polariton branches. The red dashed lines show the dispersion of the lowest two exciton states for these values of the field and the blue dotted line is the bare cavity mode. Anti-crossings of a few ${\rm meV}$ are typical for the strong-coupling regime. The solid green line shows a parabola fitted to the bottom of the polariton ground state dispersion, $E_{\rm fit}(K) = E_{\rm fit}(0) + \alpha K^2$. This is used to extract the polariton effective mass defined as $m^* = \hbar^2/(2\alpha)$. The effective mass is useful, for example, to identify the critical temperature for Bose-Einstein condensation.

\begin{figure}
\centering
\includegraphics[width=0.45\textwidth]{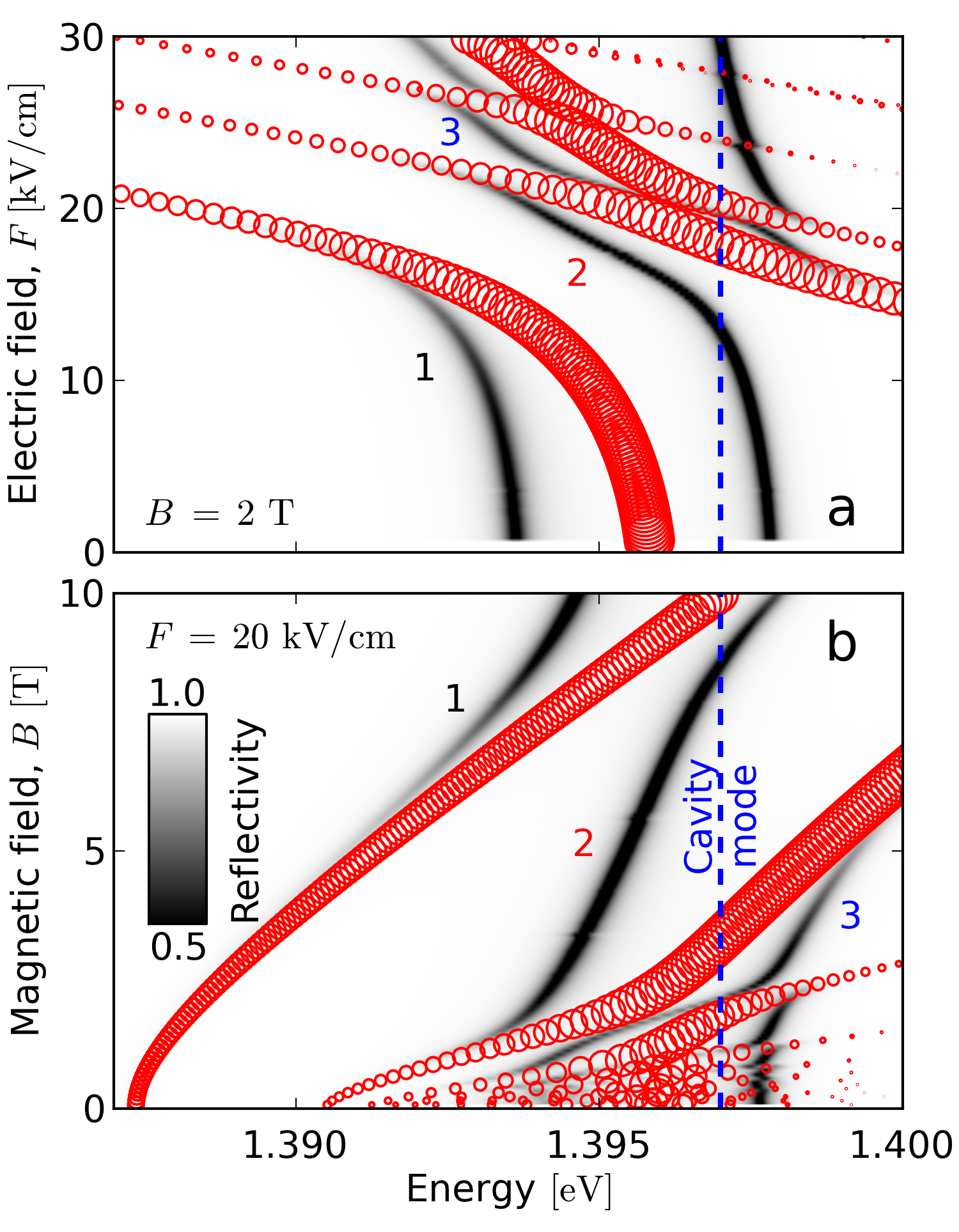}
\caption{Polariton reflectivity spectrum (grayscale) for (a) the electric field dependence with $B=2\,{\rm T}$ and (b) the magnetic field dependence with $F=20\,{\rm kV/cm}$. Exciton states are shown by circles with area proportional to the oscillator strength. The cavity mode is shown by the blue dashed line. The lowest three polariton states are labeled.}
\label{spectrum}
\end{figure}
The magneto-exciton energies are shown by red circles in Fig.\,\ref{spectrum}a as a function of electric field $F$, for magnetic field $B=2\,{\rm T}$. The circle area is proportional to the exciton oscillator strength. Bright direct excitons have an energy that changes slowly with $F$. In contrast, dark indirect excitons quickly decrease in energy with increasing $F$ due to their large dipole moment which is oriented to screen the electric field. The indirect exciton energy shift is approximately $-ed_zF$ where $d_z\approx 14\,{\rm nm}$ is the center-to-center distance of the QWs. Around $F = 15\,{\rm kV/cm}$, there is an anti-crossing between direct and indirect states. Correspondingly, there is a crossover of the ground state from a direct to an indirect exciton with increasing $F$. In Fig.\,\ref{spectrum}b, the same quantities are shown but as a function of magnetic field with $F=20\,{\rm kV/cm}$. The diamagnetic shift of the exciton ground state is seen. Also, the brightness increases with $B$. This is due to the magnetic field shrinking the in-plane part of the exciton wave function---an effect arising from tightening of the electron and hole cyclotron orbits. The shrinkage increases the overlap of electron and hole wave functions and thus enhances the coupling to light.

The grayscale in Figs.\,\ref{spectrum}a and \ref{spectrum}b, shows the corresponding dependence of the structure's reflectivity spectrum. The displacement of the polariton states from the cavity mode (shown by the blue dashed line) indicates the strength of the polariton effect. Exciton states with large oscillator strength or those that are close in energy to the cavity mode significantly modify the reflectivity. The results presented in Fig.\,\ref{spectrum} can be summarized as follows. Increasing electric field causes the direct-to-indirect crossover of the exciton ground state\Cite{SivalertpornPRB2012} and impresses a dipole moment upon the exciton part of the polariton. However, it comes at the cost of reducing the exciton brightness and eventually leads to the successive disappearance of the lowest polariton branches. The magnetic field has the reverse effect and enhances the coupling strength. It can also induce an inverse, indirect-to-direct exciton crossover which diminishes the dipole moment\Cite{WilkesNJP2016}. Both electric and magnetic field provide a means to tune the exciton energy with respect to the cavity mode.

\begin{figure}
\centering
\includegraphics[width=0.45\textwidth]{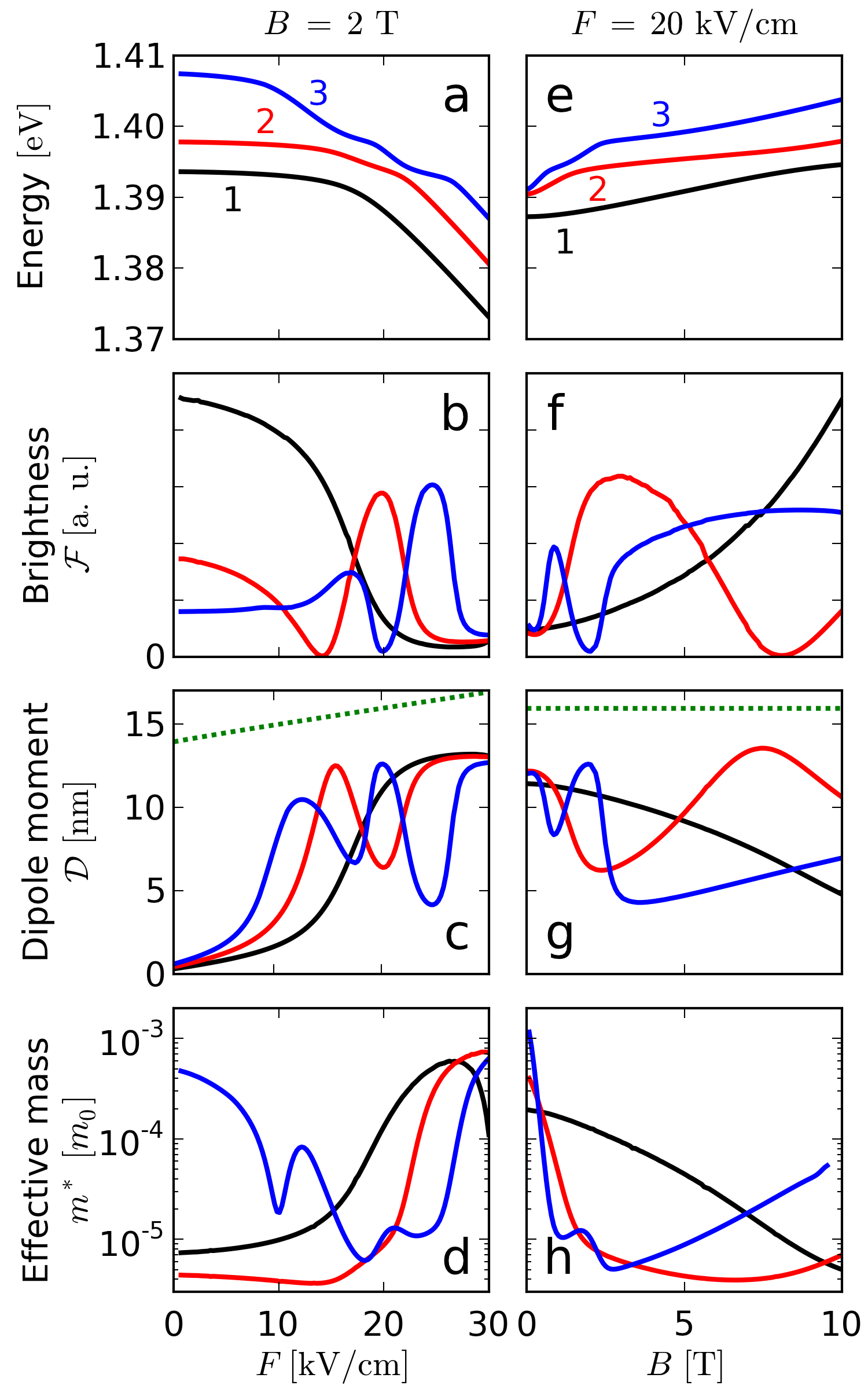}
\caption{Energy (a,e), brightness (b,f), static dipole moment (c,g) and effective mass (d,h) of the lowest three polariton states as a function of $F$ with $B = 2\,{\rm T}$ (a-d) and as a function of $B$ with $F = 20\,{\rm kV/cm}$ (e-h). The states in (a-d) and (e-h) correspond to those labeled in Fig.\,\ref{spectrum}a and \ref{spectrum}b, respectively. The dotted lines in (c) and (g) are the maximum nominal dipole moment\Cite{SivalertpornPRL2015}.}
\label{EFD}
\end{figure}

We examine the field dependencies further in Fig.\,\ref{EFD}. For the lowest three polariton states which are labeled in Figs.\,\ref{spectrum}a and \ref{spectrum}b, we identify the polariton energy as the location of the minimum in the reflectivity spectrum. These energies are plotted in Figs.\,\ref{EFD}a and \ref{EFD}e. To extract the properties of each state we evaluate the microscopic polarization $Y$ and  define the polariton brightness $\mathcal{F}$ and dipole moment $\mathcal{D}$, as in\Cite{SivalertpornPRL2015}, by
\begin{eqnarray}
\mathcal{F} &=& \frac{1}{\mathcal{N}} \left| \int Y({\bf R},0,z,z)\,dz \right|^2, \\
\mathcal{D} &=& \frac{1}{\mathcal{N}} \iiint \left| Y({\bf R},\boldsymbol\rho,z_e,z_h) \right|^2 (z_e - z_h)d\boldsymbol\rho\,dz_e\,dz_h,\,\,\,\,\,\,\,\,\,\,\, \\
\mathcal{N} &=& \iiint \left| Y({\bf R},\boldsymbol\rho,z_e,z_h) \right|^2d\boldsymbol\rho\,dz_e\,dz_h,
\end{eqnarray}
where $\mathcal{N}$ is a normalizing constant. Note that the above quantities do not depend on ${\bf R}$ due to the modulus of $Y$ (see the explicit form of $Y$ in Appendix~\ref{App0}). $\mathcal{F}$, shown in Figs.\,\ref{EFD}b and \ref{EFD}f, measures the contribution of the cavity photon and increases with the electron-hole overlap integral. $\mathcal{D}$, shown in Figs.\,\ref{EFD}c and \ref{EFD}g, is the sum of the dipole moments of all exciton states included in the summation in \Eq{chinm}, each weighted by their contribution to the total polariton wave function. The left (right) panels in Fig.\,\ref{EFD} correspond to the upper (lower) panel in Fig.\,\ref{spectrum}.

Fig.\,\ref{EFD} shows that the model predicts the existence of dipolaritons characterized by a static dipole length that can be comparable to the nominal center-to-center distance of the CQWs. Moreover, fields $F$ and $B$ provide a way to tune the dipole moments. Oscillations of the brightness (Fig.\,\ref{EFD}b and \ref{EFD}f) and dipole moment (Fig.\,\ref{EFD}c and \ref{EFD}g) are seen when changing either electric or magnetic field. Such dependence can be explained in terms of the anti-crossings between different polariton branches. Similar to indirect excitons, the electric field causes an energy shift of dipolaritons that depends linearly on the dipole moment (approximately $-e\mathcal{D}F$). Conversely, the dipole moment is approximately proportional to the energy gradient $-\partial E/\partial F$ (here $E$ is the exciton energy). The intricate anticrossings between polariton branches, which originate from anticrossings between the different direct and indirect exciton states, causes variations in this gradient (see Figs.\,\ref{spectrum}a and \ref{EFD}a). Oscillations in the dipole moment are therefore ascribed to $\partial E/\partial F$ oscillating in $F$ and $B$. In comparing Fig.\,\ref{EFD}b and \ref{EFD}c (or \ref{EFD}f and \ref{EFD}g), we see that any increase in the dipole length is accompanied by a decrease in brightness. This happens as the exciton part of the polariton transforms from a bright direct to a dark indirect state.

In Figs.\,\ref{EFD}d and \ref{EFD}h, we show the electric and magnetic field dependence of the polariton effective mass, $m^*$. This was determined by fitting a parabola to the polariton dispersion around $K=0$ as illustrated in Fig.\,\ref{schematic}c. The dependence of $m^*$ on $F$ and $B$ can be understood using a simple coupled oscillator model. The dispersion and, in turn, the mass of the polariton is approximated by diagonalizing a $2\times 2$ Hamiltonian whose diagonal elements are the dispersionless exciton and the parabolic dispersion of the cavity photon. The off-diagonal elements are a coupling constant. In this approach, one finds that $m^*$ decreases sharply with increasing coupling and that the dependence on the coupling constant is much greater when the exciton energy is below the cavity mode. This accounts for some of the main features of the lower polariton effective mass seen in Figs.\,\ref{EFD}d and \ref{EFD}h. Indeed, comparing with Figs.\,\ref{EFD}b and \ref{EFD}f, we see that an increase in brightness, which corresponds to an increase in coupling strength, sharply decreases $m^*$.

The combination of fields allows to tune the polariton mass over two orders of magnitude. Figs.\,\ref{EFD}f, \ref{EFD}g and \ref{EFD}h show that in the limit of high electric and magnetic fields ($F = 20\,{\rm kV/cm}$ and $B = 10\,{\rm T}$) the polariton ground state acquires a substantial dipole moment ($\approx 5\,{\rm nm}$) whilst remaining comparably bright and has an effective mass less than $10^{-5}\,m_0$. This particular realization of magneto-dipolaritons, whose dispersion is shown in Fig.\,\ref{schematic}c, offers favorable conditions for low condensation threshold and strong polariton-polariton interactions that can be used to probe many-body interactions.

\begin{figure}
\centering
\includegraphics[width=0.45\textwidth]{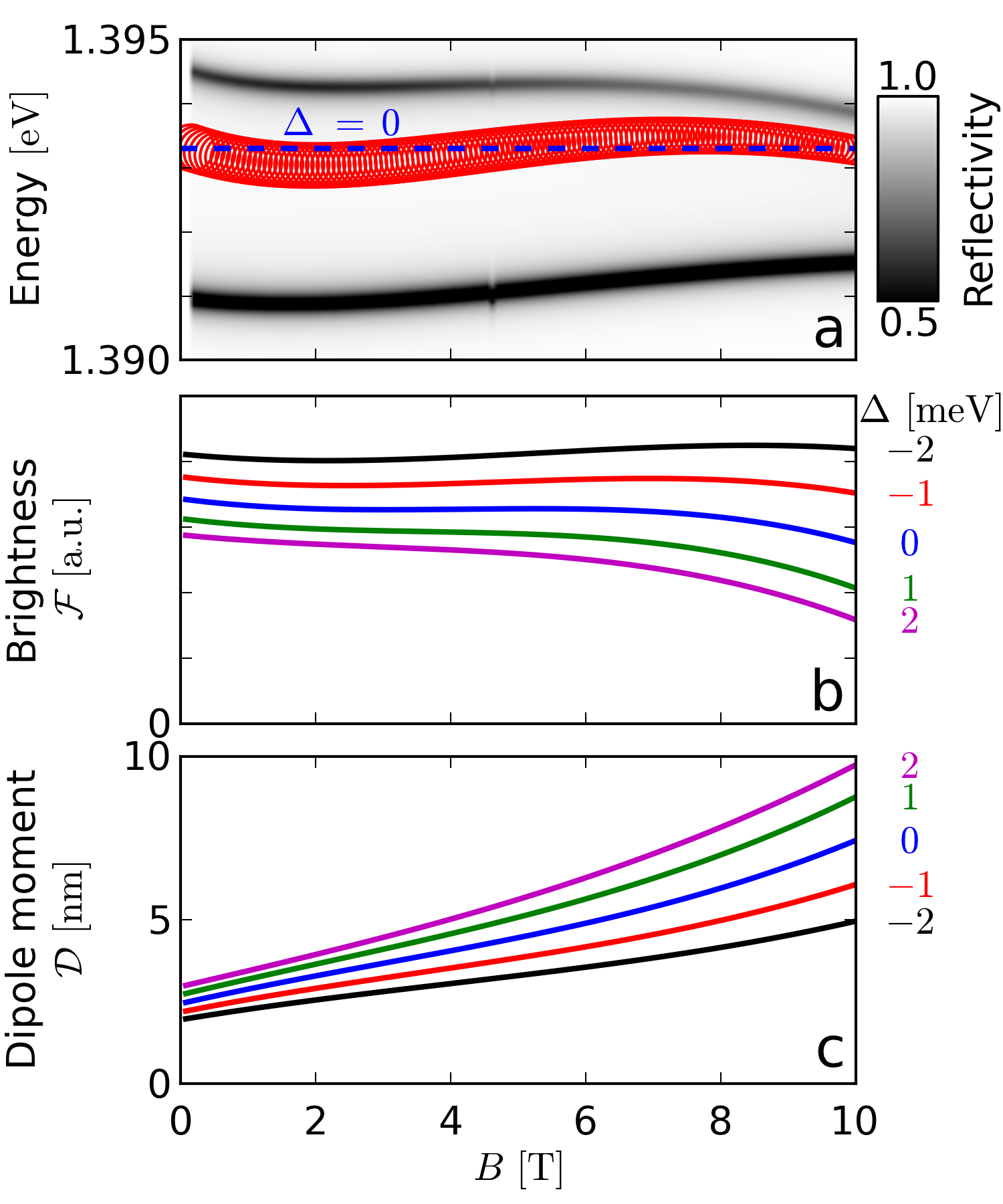}
\caption{(a) Reflectivity spectrum (grayscale), exciton energy (red circles with area proportional to oscillator strength) and cavity mode for zero detuning (blue dashed line). Brightness (b) and dipole moment (c) of the polariton ground state for various detunings of the cavity mode from the exciton ground state, as given.}
\label{fixedEnergy}
\end{figure}
We now explore the properties of magneto-dipolaritons in a fixed energy scheme. $F$ and $B$ are increased simultaneously so that the diamagnetic shift of the exciton ground state is compensated for by the electric field. This is approximately achieved with an electric field that is linear in $B$, going from $F=13\,{\rm kV/cm}$ at $B=0$ up to $F=23\,{\rm kV/cm}$ at $B=10\,{\rm T}$. The resulting energy of the exciton ground state is shown by red circles in Fig.\,\ref{fixedEnergy}a and is practically constant. We then used cavity modes with different detunings $\Delta$ from the exciton ground state. The grayscale in Fig.\,\ref{fixedEnergy}a shows the reflectivity spectrum for the case of $\Delta=0$ and the cavity mode is shown by the blue dashed line.

Increasing the electric field reduces the spatial overlap of electron and hole wave functions and darkens the polariton state. In the fixed energy scheme, this is compensated for by the magnetic field which shrinks the radius of the exciton wave function in the QW plane. The brightness of the polariton ground state is shown in Fig.\,\ref{fixedEnergy}b. We see only a weak dependence on the fields. In contrast, the dipole moment of the polariton ground state, shown in Fig.\,\ref{fixedEnergy}c increases substantially due to the increasing electric field. For example, with $\Delta=0$ the dipole moment is increased three fold at the rather minor sacrifice of about a $15\%$ decrease in brightness. The magnetic field can therefore be used to create polaritons that are both bright and have a large static dipole moment.

\section{Conclusion}
\label{summarySec}
We implemented a precise calculation of exciton-polariton states in microcavity embedded CQWs in the presence of external electric and magnetic fields oriented perpendicular to the QW plane. We evaluated the energy, brightness, dipole moment and effective mass of the lowest three polariton states. We found a trade-off between the brightness and dipole moment as increasing the latter reduces the exciton oscillator strength which in turn depends on the electron-hole overlap. However, the reduction of the polariton brightness caused by an electric field-induced dipole moment can be compensated by a magnetic field which shrinks the exciton wave function in the QW plane. As as result, a polariton with a substantial dipole moment can be formed in the strong coupling regime, also showing a significantly reduced (down to $10^{-5}m_0$) polariton effective mass.

\section*{Acknowledgments}
Support of this work by the EPSRC (grant EP/L022990/1) is gratefully acknowledged. Computational work was done using the facilities of the ARCCA Division, Cardiff University.

\appendix

\section{Excitonic susceptibility}
\label{App0}
Here we derive the expression for the excitonic susceptibility \Eq{susceptibility}, using the spectral representation of the exciton Green's function (GF) $G({\bf R},\boldsymbol\rho,z_e,z_h;{\bf R}',\boldsymbol\rho',z_e',z_h')$. The latter satisfies an equation
\begin{equation}
(\hat{H}-\hbar\omega-i\gamma) G=\delta({\bf R}-{\bf R}') \delta(\boldsymbol\rho-\boldsymbol\rho')\delta(z_e-z_e')\delta(z_h-z_h').
\end{equation}
Then the solution of the material equation (\ref{materialEqn}) can be expressed as a convolution with the GF
\begin{eqnarray}
&Y({\bf R},\boldsymbol\rho,z_e,z_h)=
\int\,d{\bf r}'\int d{\bf R}'\int dz' \mathbfcal{M}({\bf r}') \cdot {\bf E}_\omega({\bf R}',z')\nonumber
\\
&\times G({\bf R},\boldsymbol\rho,z_e,z_h;{\bf R}',\boldsymbol\rho',z_e',z_h')\,.
\end{eqnarray}
Using an isotropic distribution in the QW plane of the transition dipole moment, we note that $\mathbfcal{M}$ is proportional to the polarization vector $\hat{\bf e}$. Then having $\mathbfcal{M}({\bf r})= \hat{\bf e} \mu_{\rm cv}\delta({\bf r})$ in the point dipole approximation and taking the light field in the form of \Eq{lightField}, the microscopic polarization reduces to
\begin{eqnarray}
Y({\bf R},\boldsymbol\rho,z_e,z_h)&=&\mu_{\rm cv} \int d{\bf R}'\int dz' e^{i{\bf K}\cdot {\bf R}'} {\cal E}(z')
\nonumber
\\
&&\times G({\bf R},\boldsymbol\rho,z_e,z_h;{\bf R}',0,z',z').
\end{eqnarray}
Now, using the spectral representation of the GF
\begin{eqnarray}
G&=&\frac{1}{S}\sum_{\bf K} g_{\bf K}({\bf R},\boldsymbol\rho) g^\ast_{\bf K}({\bf R}',\boldsymbol\rho')
\nonumber
\\
&&\times\sum_\nu \frac{\varphi_{\nu{\bf K}}(\boldsymbol\rho,z_e,z_h) \varphi_{\nu{\bf K}}^\ast(\boldsymbol\rho',z_e',z_h')}{E_{\nu{\bf K}} - \hbar\omega- i \gamma},
\label{GFSR}
\end{eqnarray}
where
\begin{equation}
g_{\bf K}({\bf R},\boldsymbol\rho)=\exp \left(i\left[{\bf K} + \frac{e}{\hbar c}{\bf A}(\boldsymbol\rho) \right] \cdot {\bf R}\right)
\end{equation}
and $S$ is the QW normalization area, obtain
\begin{eqnarray}
Y({\bf R},\boldsymbol\rho,z_e,z_h)&=&\mu_{\rm cv} e^{i{\bf K}\cdot{\bf R}} \sum_\nu \frac{\varphi_{\nu{\bf K}}(\boldsymbol\rho,z_e,z_h) }{E_{\nu{\bf K}} - \hbar\omega- i \gamma}
\nonumber
\\
&&\times\int\varphi_{\nu{\bf K}}^\ast(0,z',z') {\cal E}(z')dz'.
\end{eqnarray}
Using the definition of the excitonic susceptibility \Eq{susc}
then obtain
\begin{equation}
\chi_\omega(z,z')= \mu_{\rm cv}^2 \sum_\nu \frac{\varphi_{\nu{\bf K}}(0,z,z) \varphi_{\nu{\bf K}}^\ast(0,z',z')}{E_{\nu{\bf K}} - \hbar\omega- i \gamma}\,,
\end{equation}
where the dependence on light frequency is explicitly indicated. This series summation, as well as the representation \Eq{GFSR} of the exciton GF itself, has known convergence issues\Cite{Zimm86} in 2D and 3D, which can be eliminated by subtracting the zero frequency value, already included into the static dielectric constant $\epsilon_b(z)$. The susceptibility is then redefined as
\begin{equation}
\chi(z,z')=\chi_\omega(z,z')-\chi_0(z,z')
\end{equation}
leading to the expression given in \Eq{susceptibility}.

\section{Scattering matrix for the active CQW layers}
\label{appA}

We first simplify the wave equation \Eq{wave}, by introducing $Q(z,z')$ defined as
\begin{equation}
Q(z,z') = -4\pi\frac{\omega^2}{c^2}\chi(z,z') = \sum_{n,m}\Phi_n(z,z)\Phi_m(z',z')Q_{nm},
\end{equation}
where
\begin{equation}
Q_{nm} = -4\pi\frac{\omega^2}{c^2} \chi_{nm}
\label{QnmEqn}
\end{equation}
with $ \chi_{nm}$ given by \Eq{chinm}. Maxwell's wave equation then takes the form
\begin{equation}
\left(\frac{\partial^2}{\partial z^2} + q^2\right)\mathcal{E}(z) = \int_{-\infty}^{\infty}Q(z,z')\mathcal{E}(z')\,dz',
\label{newMaxwell}
\end{equation}
where $q^2 = \epsilon_b \omega^2/c^2 - K^2$. \Eq{newMaxwell} can be solved with the help of the free space GF $G(z,z')$ that satisfies
\begin{equation}
\left(\frac{\partial^2}{\partial z^2} + q^2\right)G(z,z') = \delta(z-z')
\label{greenEqn}
\end{equation}
which, with outgoing boundary conditions, has the following form:
\begin{equation}
G(z,z') = \frac{e^{iq|z-z'|}}{2iq}.
\label{greensFunc}
\end{equation}
Then, the solution to \Eq{newMaxwell} satisfies an integral equation,
\begin{eqnarray}
\mathcal{E}(z)& =& A^+e^{iqz} + A^-e^{-iqz}\label{integralEqn}\\
&&+ \int_{-\infty}^{\infty}dz'\int_{-\infty}^{\infty}dz''\,G(z,z')Q(z',z'')\mathcal{E}(z'').\nonumber
\end{eqnarray}
Next, we introduce
\begin{eqnarray}
X_n = \int_{-\infty}^{\infty}\mathcal{E}(z)\Phi_n(z,z)\,dz, 
\end{eqnarray}
so that \Eq{integralEqn} becomes
\begin{eqnarray}
\mathcal{E}(z) &=& A^+e^{iqz} + A^-e^{-iqz}\label{newIntegralEqn}\\
&&+ \sum_{n,m}\int_{-\infty}^{\infty}dz'\,G(z,z')\Phi_n(z',z')Q_{nm}X_m.
\nonumber
\end{eqnarray}
After multiplying \Eq{newIntegralEqn} by $\Phi_n(z,z)$ and integrating over $z$, one gets
\begin{equation}
X_n = A^+\tilde{\Phi}_n(q) + A^-\tilde{\Phi}_n(-q) + \sum_{m,l}G_{nm}Q_{ml}X_l,
\label{Xn}
\end{equation}
where
\begin{eqnarray}
\tilde{\Phi}_n(q) = \int_{-\infty}^{\infty}e^{iqz}\Phi_n(z,z)\,dz
\label{PhiTilde}
\end{eqnarray}
and 
\begin{equation}
G_{nm} = \int_{-\infty}^{\infty}dz\int_{-\infty}^{\infty}dz'\,G(z,z')\Phi_n(z,z)\Phi_m(z',z').
\label{GnmEqn}
\end{equation}
By introducing the matrix $V$, given by
\begin{equation}
V_{nm} = \sum_lG_{nl}Q_{lm},
\end{equation}
\Eq{Xn} reduces to the following linear system 
\begin{equation}
\sum_m(\delta_{nm} - V_{nm})X_m = A^+\tilde{\Phi}_n(q) + A^-\tilde{\Phi}_n(-q).
\label{linSys}
\end{equation}
\Eq{linSys} is solved for the vector $X$, using  specific boundary conditions imposed by $A^+$ and $A^-$. 
We first consider the case of an incident wave traveling in the positive $z$-direction ($A^+ = 1$ and $A^- = 0$) and find, 
using the asymptotics at $z\to\pm\infty$ of the last term in \Eq{newIntegralEqn},
\begin{equation}
\int_{-\infty}^{\infty}dz'\,G(z,z')\Phi_n(z',z')=\frac{e^{\pm iqz}\tilde{\Phi}_n(\mp q)}{2iq}\,
\end{equation}
the transmission and reflection coefficients, $T_+$ and $R_+$, respectively. These are
\begin{eqnarray}
T_+ &=& \frac{1}{2iq}\sum_{n,m} \tilde{\Phi}_n(-q)Q_{nm}X^+_m + 1, \label{TEqn} \\
R_+ &=& \frac{1}{2iq}\sum_{n,m} \tilde{\Phi}_n(+q)Q_{nm}X^+_m, \label{REqn}
\end{eqnarray}
where, $X^+_n$ satisfies
\begin{equation}
\sum_m(\delta_{nm} - V_{nm})X^+_m = \tilde{\Phi}_n(q).
\label{XnEqn}
\end{equation}
Using $A^+ = 0$ and $A^- = 1$, we then treat a wave traveling in the negative $z$-direction and find the transmission and reflection coefficients, $T_-$ and $R_-$, respectively. These are
\begin{eqnarray}
T_- &=& \frac{1}{2iq}\sum_{n,m} \tilde{\Phi}_n(+q)Q_{nm}X^-_m + 1, \label{tEqn} \\
R_- &=& \frac{1}{2iq}\sum_{n,m} \tilde{\Phi}_n(-q)Q_{nm}X^-_m, \label{rEqn}
\end{eqnarray}
where, $X^-_n$ satisfies
\begin{equation}
\sum_m(\delta_{nm} - V_{nm})X^-_m = \tilde{\Phi}_n(-q).
\label{YnEqn}
\end{equation}
The scattering matrix of the active region is then constructed as
\begin{eqnarray}
S^{({\rm CQW})} = \left( \begin{array}{cc}
T_+ & R_- \\
R_+ & T_- \end{array} \right).
\end{eqnarray}
In conjunction with the scattering matrices of each layer of the DBRs and the cavity, the total scattering matrix, which describes the optical response of the entire structure, is formed following Ref.\Cite{KoPRB1988}.

\section{Computational algorithm}
\label{algorithmSec}
Here, we summarize the technical implementation of the calculation of $S^{({\rm CQW})}$. Prior calculation of the basis of pair states $\Phi_n(z_e,z_h)$ and the in-plane exciton wave functions $\phi^{(\nu)}_n(\rho)$ is assumed (see Refs.\Cite{SivalertpornPRB2012,WilkesNJP2016} for details).

(i) For each electric field, the matrix $G_{nm}$ is calculated via \Eq{GnmEqn}. Due to the form of the GF \Eq{greensFunc}, which depends on $z - z'$ only, the integral is a convolution and may be optimized using fast Fourier transform.

(ii) The Fourier transforms $\tilde{\Phi}_n(\pm q)$ in \Eq{PhiTilde} are evaluated.

(iii) For each photon energy and electric and magnetic field, the matrix $Q_{nm}$ is calculated using Eqs.\,(\ref{QnmEqn}) and (\ref{chinm}). Here, a sufficiently large number of exciton states must be included so that the summation converges with respect to $\nu$. In practice, up to 200 states may be required. Calculating these states is the main computational cost of the process.

(iv) Solve the linear systems for $X^+_n$ and $X^-_n$ given by \Eq{XnEqn} and \Eq{YnEqn}, respectively and evaluate the matrix $S^{({\rm CQW})}$ via Eqs.~(\ref{TEqn}), (\ref{REqn}), (\ref{tEqn}), and (\ref{rEqn}).


\begin{thebibliography}{99}

\bibitem{CarusottoRMP2013}
I. Carusotto and C. Ciuti, Rev. Mod. Phys. {\bf 85}, 299 (2013).

\bibitem{KasprzakNAT2006}
J. Kasprzak, M. Richard, S. Kundermann, A. Baas, P. Jeambrun, J. M. J. Keeling, F. M. Marchetti, M. H. Szymanska, R. Andre, J. L. Staehli, V. Savona, P. B. Littlewood, B. Deveaud, and L. S. Dang, Nature {\bf 443}, 409 (2006).

\bibitem{BaliliSCI2007}
R. Balili, V. Hartwell, D. Snoke, L. Pfeiffer and K. West, Science {\bf 316}, 1007 (2007).

\bibitem{AmoNAT2009}
A. Amo, D. Sanvitto, F. P. Laussy, D. Ballarini, E. d. Valle, M. D. Martin, A. Lemaitre, J. Bloch, D. N. Krizhanovskii, M. S. Skolnick, C. Tejedor and L. Vina, Nature {\bf 457}, 291 (2009).

\bibitem{LagoudakisNATP2008}
K. G. Lagoudakis, M. Wouters, M. Richard, A. Baas, I. Carusotto, R. Andre, L. S. Dang and B. Deveaud-Pledran, Nat. Phys. {\bf 4}, 706 (2008).

\bibitem{ChristopoulosPRL2007}
S. Christopoulos, G. Baldassarri Höger von Högersthal, A. J. D. Grundy, P. G. Lagoudakis, A. V. Kavokin, J. J. Baumberg, G. Christmann, R. Butte, E. Feltin, J.-F. Carlin, and N. Grandjean, Phys. Rev. Lett. {\bf 98}, 126405 (2007).

\bibitem{BhattacharyaPRL2014}
P. Bhattacharya, T. Frost, S. Deshpande, M. Z. Baten, A. Hazari and A. Das, Phys. Rev. Lett. {\bf 112}, 236802 (2014).

\bibitem{BallariniNATC2013}
D. Ballarini, M. De. Giorgi, E. Cancellieri, R. Houdre, E. Giacobino, R. Cingolani, A. Bramati, G. Gigli and D.
Sanvitto, Nat. Commun. {\bf 4}, 1778 (2013).

\bibitem{ButovNAT2002}
L. V. Butov, A. C. Gossard and D. S. Chemla, Nature {\bf 418}, 751 (2002).

\bibitem{ButovJETP2016}
L. V. Butov, JETP {\bf 149}, 505 (2016).

\bibitem{ButovPRB1999}
L. V. Butov, A. Imamoglu, A. V. Mintsev, K. L. Campman and A. C. Gossard, Phys. Rev. B {\bf 59}, 1625 (1999).

\bibitem{NegoitaPRB2000}
V. Negoita, D. W. Snoke and K. Eberl, Phys. Rev. B {\bf 61}, 2779 (2000).

\bibitem{LaikhtmanEPL2009}
B. Laikhtman and R. Rapaport, Europhys. Lett. {\bf 87}, 27010 (2009).

\bibitem{Cohen_arXiv2015}
K. Cohen, Y. Shilo, R. Rapaport, K. West and L. Pfeiffer, arXiv:1506.09058 (2015).

\bibitem{WinbowPRL2011}
A. G. Winbow, J. R. Leonard, M. Remeika, Y. Y. Kuznetsova, A. A. High, A. T. Hammack, L. V. Butov, J. Wilkes, A.A. Guenther, A. L. Ivanov, M. Hanson and A. C. Gossard, Phys. Rev. Lett. {\bf 106}, 196806 (2011).

\bibitem{LeonardAPL2012}
J. R. Leonard, M. Remeika, M. K. Chu, Y. Y. Kuznetsova, A. A. High, L. V. Butov, J. Wilkes, M. Hanson and A. C. Gossard, Appl. Phys. Lett. {\bf 100}, 231106 (2012).

\bibitem{DorowAPL2016}
C. J. Dorow, Y. Y. Kuznetsova, J. R. Leonard, M. K. Chu, L. V. Butov, J. Wilkes, M. Hanson and A. C. Gossard, Appl. Phys. Lett. {\bf 108}, 073502 (2016).

\bibitem{SchinnerPRL2013}
G. J. Schinner, J. Repp, E. Schubert, A. K. Rai, D. Reuter, A. D. Wieck, A. O. Govorov, A. W. Holleitner and J. P. Kotthaus, Phys. Rev. Lett. {\bf 110}, 127403 (2013).

\bibitem{AlloingSREP2013}
M. Alloing, A. Lemaitre, E. Galopin and F. Dubin, Sci. Rep. {\bf 3}, 1578 (2013).

\bibitem{HammackPRL2006}
A. T. Hammack, M. Griswold, L. V. Butov, L. E. Smallwood, A. L. Ivanov and A. C. Gossard, Phys. Rev. Lett. {\bf 96}, 227402 (2006).

\bibitem{AndreakouAPL2014}
P. Andreakou, S. V. Poltavtsev, J. R. Leonard, E. V. Calman, M. Remeika, Y. Y. Kuznetsova, L. V. Butov, J. Wilkes, M. Hanson and A.C. Gossard, Appl. Phys. Lett. {\bf 104}, 091101 (2014).

\bibitem{CristofoliniSCI2012}
P. Cristofolini, G. Christmann, S. I. Tsintzos, G. Deligeorgis, G. Konstantinidis, Z. Hatzopoulos, P. G. Savvidis and J. J. Baumberg, Science {\bf 336}, 704 (2012).

\bibitem{KyriienkoPRL2013}
O. Kyriienko, A. V. Kavokin and I. A. Shelykh, Phys. Rev. Lett. {\bf 111}, 176401 (2013).

\bibitem{KyriienkoPRA2014}
O. Kyriienko, I. A. Shelykh, and T. C. H. Liew, Phys. Rev. A {\bf 90}, 033807 (2014).

\bibitem{KhadzhiJETP2015}
P. I. Khadzhi and O. F. Vasilieva, JETP Lett. {\bf 102}, 665 (2015).

\bibitem{Rosenberg_arXiv2016}
I. Rosenberg, Y. Harpaz, K. West, L. Pffeifer and R. Rapaport, arXiv:1604.05952 (2016).

\bibitem{TignonPRL1995}
J. Tignon, P. Voisin, C. Delalande, M. Voos, R. Houdre, U. Oesterle and R. P. Stanley, Phys. Rev. Lett. {\bf 74}, 3967 (1995).

\bibitem{FisherPRB1996}
T. A. Fisher, A. M. Afshar, M. S. Skolnick, D. M. Whittaker and J. S. Roberts, Phys. Rev. B {\bf 53}, R10469 (1996).

\bibitem{PietkaPRB2015}
B. Pietka, D. Zygmunt, M. Krol, M. R. Molas, A. A. L. Nicolet, F. Morier-Genoud, J. Szczytko, J. Lusakowski, P. Zieba, I. Tralle, P. Stepnicki, M. Matuszewski, M. Potemski, and B. Deveaud, Phys. Rev. B {\bf 91}, 075309 (2015).

\bibitem{BhattacharyaAPL2012}
P. Bhattacharya, A. Das, S. Bhowmick, M. Jankowski and C. Lee, Appl. Phys. Lett. {\bf 100}, 171106 (2012).

\bibitem{Kochereshko_arXiv2013}
V. P. Kochereshko, M. V. Durnev, L. Besombes, H. Mariette, V. F. Sapega, A. Axitopoulos, I. G. Savenko, T. C. H. Liew, I. A. Shelykh, A. V. Platonov, S. I. Tsintzos, Z. Hatzopoulos, P. Lagoudakis, P. G. Savvidis, C. Schneider, M. Amthor, C. Metzger, M. Kamp, S. Hoefling and A. Kavokin, arXiv:1309.6983 (2013).

\bibitem{SivalertpornPRB2012}
K. Sivalertporn, L. Mouchliadis, A. L. Ivanov, R. Philp and E. A. Muljarov, Phys. Rev. B {\bf 85}, 045207 (2012).

\bibitem{WilkesNJP2016}
J. Wilkes and E. A. Muljarov, New J. Phys. {\bf 18}, 023032 (2016).

\bibitem{LozovikJETP1997}
Yu E. Lozovik and A. M. Ruvinskii, JETP {\bf 85}, 979 (1997).

\bibitem{ArseevJETP1998}
P. I. Arseev and A. B. Dzyubenko, JETP {\bf 87}, 200 (1998).
%\bibitem{DzyubenkoPRB1996}
%A. B. Dzyubenko and A. L. Yablonskii, Phys. Rev. B {\bf 53}, 16355 (1996).

\bibitem{SivalertpornPLA2016}
K. Sivalertporn, Phys. Lett. A {\bf 380}, 1990 (2016).

\bibitem{SivalertpornPRL2015}
K. Sivalertporn and E. A. Muljarov, Phys. Rev. Lett. {\bf 115}, 077401 (2015).

\bibitem{Stahl87} A. Stahl and I. Balslev, {\em Electrodynamics of the Semiconductor Band Edge} (Springer-Verlag, Berlin, 1987).

\bibitem{KoPRB1988}
D. Ko and J. C. Inkson, Phys. Rev. B {\bf 38}, 9945 (1988).

\bibitem{Zimm86} R. Zimmermann, Phys. Stat. Sol. (b) {\bf 135}, 681 (1986).

\end{thebibliography}
\end{document}